%% file: main.tex
\documentclass[sigconf,natbib=false, 9pt]{acmart}
\AtBeginDocument{%
  }
\RequirePackage[
  datamodel=acmdatamodel,
  sorting=none, 
  style=nature,
  ]{biblatex}
\usepackage[T1]{fontenc}
\usepackage{url}

\usepackage{breakurl}
\usepackage{hyperref}
\usepackage{ae,aecompl}
\usepackage{times}
\usepackage{lipsum}
\usepackage{graphicx}
\usepackage{subfig}
\usepackage{epsfig}
\usepackage{epstopdf}
\usepackage{wrapfig}
\usepackage{longtable} 
\usepackage{threeparttable}
\usepackage{amsmath}
\usepackage{float}
\usepackage{soul}

\usepackage{multirow}
\usepackage{subfig}
\usepackage{tabu}
\usepackage{tabularx}
\usepackage{booktabs}

\usepackage{rotating}
\usepackage{titlesec}
\usepackage{hyperref}
\usepackage[normalem]{ulem}
\usepackage{caption}
\usepackage{subcaption}
\usepackage{ccicons}

\usepackage{enumitem}
\DeclareCaptionType{copyrightbox}
\makeatletter
\def\@copyrightspace{\relax}
\makeatother

\copyrightyear{2024} 
\acmYear{2024} 
\setcopyright{rightsretained} 
\acmConference[DAC '24]{61st ACM/IEEE Design Automation Conference}{June 23--27, 2024}{San Francisco, CA, USA}
\acmBooktitle{61st ACM/IEEE Design Automation Conference (DAC '24), June 23--27, 2024, San Francisco, CA, USA}
\acmDOI{10.1145/3649329.3656531}
\acmISBN{979-8-4007-0601-1/24/06}

\begin{document}

\title{Uncovering Software-Based Power Side-Channel Attacks on Apple M1/M2 Systems}

\author{Nikhil Chawla, Chen Liu, Abhishek Chakraborty, Igor Chervatyuk, Ke Sun, Thaís Moreira Hamasaki, Henrique Kawakami}

\email{{nikhil.chawla, chen1.liu, abhishek1.chakraborty, igor.chervatyuk, ke.sun, thais.moreira.hamasaki, henrique.kawakami}@intel.com}
\affiliation{
\institution{Intel Corporation, USA}
  \country{}
}

\renewcommand{\shortauthors}{N. Chawla, C. Liu, A. Chakraborty, I. Chervatyuk, K. Sun, T. Moreira Hamasaki, H. Kawakami}

\begin{abstract}
Traditionally, power side-channel analysis requires physical access to the target device, as well as specialized devices to measure the power consumption with enough precision.
Recently research has shown that on x86 platforms, on-chip power meter capabilities exposed to a software interface might be used for power side-channel attacks without physical access. In this paper, we show that such software-based power side-channel attack is also applicable on Apple silicon (e.g., M1/M2 platforms), exploiting the System Management Controller (SMC) and its power-related keys, which provides access to the on-chip power meters through a software interface to user space software. 
We observed data-dependent power consumption reporting from such SMC keys and analyzed the correlations between the power consumption and the processed data. Our work also demonstrated how an unprivileged user mode application successfully recovers bytes from an AES encryption key from a cryptographic service supported by a kernel mode driver in MacOS. We have also studied the feasibility of performing frequency throttling side-channel attack on Apple silicon. Furthermore, we discuss the impact of software-based power side-channels in the industry, possible countermeasures, and the overall implications of software interfaces for modern on-chip power management systems. 
\end{abstract}

\maketitle
\section{Introduction}\label{s:intro}
\input{sections/intro}
\section{Background \& Related Work}\label{s:related}
\input{sections/related}
\section{SMC power meter side-channel attacks}\label{s:attack}
\input{sections/method}
\section{Throttling side-channel analysis on M2}\label{s:throttling}

\input{sections/throttling}

\section{Countermeasures \& Discussions}\label{s:mitigation}
\input{sections/mitigation}

\section{Conclusion}\label{s:conclusion}
\input{sections/conclusion}
\AtNextBibliography{\small}
\printbibliography
\end{document}

%% file: sections/intro.tex
It is well-known that CMOS circuits, when processing data, generate data-dependent power consumption. This behavior has been misused to perform power analysis attacks, which extracts sensitive data, such as secret keys or passwords, from the target system. 
Traditionally, performing power side-channel attacks requires the attacker to have physical access to the system for power consumption measurement.
Software-based power side-channel attacks represent a new class of power side-channel attacks that can be performed by a software attacker leveraging on-chip power meter or power management capabilities provided by the hardware, without requiring physical access to the system. One example of such attacks is the \emph{PLATYPUS} \cite{Lipp2020Platypus}, which leverages the Running Average Power Limit (RAPL) energy counters available on x86 CPUs to extract sensitive information. Another example is frequency throttling attack \cite{wan2022hertzbleed, 10.1145/3548606.3560682}, which further converts the power side-channel to frequency/timing side-channel, allows information leakage even without exposing power meter interfaces. 

While most of these researches are targeting x86 CPUs, researches on applicability of such attacks targeting other architectures is limited. In this paper, for the first time, we perform a comprehensive analysis of software-based power side-channel attack targeting ARM-based Apple M1/M2 systems. Specifically, we observed that the System Management Controller (SMC) on Apple silicon exposes power meter capabilities to software, and a set of such metrics are even accessible to user-space application in macOS. With experiments, we identified the power meters that are correlated with the data processed by the CPU. Furthermore, we show that the data-dependency can be exploited by user-space software that perform power side-channel attacks to extract secrets (e.g., AES \cite{AES} keys) from a privileged software (e.g., kernel drivers). Besides that, we have also studied on an alternative power reporting interface and the feasibility of performing throttling side-channel attacks on Apple silicon. 
Our key contributions in the paper are:

\begin{itemize}
    \item To the best of our knowledge, this is the first work that presents comprehensive analysis on feasibility of software-based power side-channel attacks on ARM-based Apple silicon, including both power-meter-based attack \cite{Lipp2020Platypus} and frequency throttling attack \cite{wan2022hertzbleed, 10.1145/3548606.3560682}.
    \item We discover that specific power meters provided by the SMCs on the Apple silicon are data-dependent and prove that with experiments. This is the first work that shows the existence of such side-channel. Moreover, we show that those power meters, which are exposed to a user mode application, allow an unprivileged attacker to infer secret (e.g., AES key) from kernel mode. Our PoC demonstrates complete leakage of 6 key bytes and nearly-complete leakage of the other 6 key bytes for a 16-byte AES-128 key. 
    \item We discuss the impact of software-based power side-channels on the industry as a whole and possible countermeasure techniques to mitigate the risks.
    
\end{itemize}

The remainder of this paper starts by presenting background and related work in Section \ref{s:related}. In Section \ref{s:attack}, we present the details of our research on finding the data-dependency of the SMC power meters reporting, as well as the exploitation of the power side-channel to extract secret. We further discuss the experiments on throttling side-channel feasibility in Section \ref{s:throttling} and discuss the possible countermeasure techniques in Section \ref{s:mitigation}. Section \ref{s:conclusion} concludes our paper.

\noindent\textbf{Responsible Disclosure:} We responsibly disclosed our findings to Apple Inc. on November 29th, 2022. Following up on Apple's request, we provided two different Proof-of-Concept (PoC) versions on January 2023 and March 2023, respectively. Apple acknowledged the findings,  validated the PoC on June 27, 2023 and did not propose to take further actions to mitigate the reported issue. 

%% file: sections/related.tex
\subsection{Power meter reporting on Apple platforms}
The System Management Controller (SMC) is a co-processor responsible for power and thermal management available in multiple Apple platforms since the legacy x86-based Apple systems\cite{InversePathSMC}. 
The SMC exposes various sensor data including temperature, voltage and power meters, battery status, fan status, and other power-related functions on the system. The sensor data is accessible to user mode software as a key-value pair, where the key is a 4-character alphanumeric string\cite{AppleSMCReCon} and the value can be retrieved by calling the \texttt{IOConnectCallStructMethod} function in the \texttt{IOKit}\cite{IOKit} library built in macOS. 

\subsection{Software-based power side-channel attacks}\label{ssec:sw-based psc}
In a traditional physical power side-channel attacks, the attacker measures the data-dependent power consumption of cryptographic implementations processing sensitive data and perform statistic-based analysis to recover the secret \cite{mangard2008power}. Correlation Power Analysis (CPA) is one of the traditional power side-channel analysis techniques that computes correlation coefficient between the actual secret dependent  power traces and hypothetical power model to recover to secret key. 

Unlike the traditional physical side-channel attacks, software-based power side-channel attacks do not require physical access and can be performed remotely by exploiting information provided by underlying hardware through software accessible interfaces.
The \emph{PLATYPUS} attack \cite{Lipp2020Platypus} has shown that the Running Average Power Limit (RAPL) energy reporting interfaces, which are available on Intel and AMD processors, expose data-dependent power consumption information that can be used to perform power side-channel attack by a software attacker. Furthermore, recent works \cite{wan2022hertzbleed, 10.1145/3548606.3560682} have discovered that CPUs, when hit certain reactive limits (e.g, power or current limits), will throttle to run at lower frequencies, which are correlated with energy consumption of the chip and hence correlated with data processed by the CPU. \emph{Collide+Power} \cite{Kogler2023CollidePower} further shows that data-dependent power consumption information from the memory hierarchy can be exposed through those channels.
Taneja \emph{et al.} \cite{taneja2023hot} discusses thermal-limit-induced frequency throttling side-channel on various integrated and discrete GPUs, and also mentions software-based power side-channels on ARM-based CPUs, including Apple M1/M2 systems. Compared to \cite{taneja2023hot}, this work conducts a more comprehensive study on the software-based power side channel on Apple M1/M2 platforms, and more importantly, finds new software-accessible interfaces that exhibits more apparent data-dependent side-channel leakage. 

%% file: sections/method.tex
In this section, we provide detailed descriptions of the procedures used to identify data-dependent SMC key values and evaluate the feasibility of performing software-based side-channel attacking using the keys. We used an Apple Mac Mini M1 and an Apple MacBook Air M2 for all the experiments described in this paper, and their specifications are summarized at Table \ref{tab:system config}.

\begin{table}[]
\footnotesize
\caption{Specifications of the tested devices}
\label{tab:system config}
\begin{tabular}{c|cc|cc|c}
\multirow{2}{*}{Devices} & \multicolumn{2}{c|}{P-cores}           & \multicolumn{2}{c|}{E-cores}           & \multirow{2}{*}{OS version} \\ \cline{2-5}
                         & \multicolumn{1}{c|}{count} & max freq. & \multicolumn{1}{c|}{count} & max freq. &                             \\ \hline
Mac Mini M1              & \multicolumn{1}{c|}{4}     & 3.2 GHz   & \multicolumn{1}{c|}{4}     &   2.4GHz        & macOS 12.5                  \\
Mac Air M2               & \multicolumn{1}{c|}{4}     & 3.5 GHz   & \multicolumn{1}{c|}{4}     &      2.06GHz     & macOS 13.0                  \\ \hline
\end{tabular}%
\end{table}

\begin{table}
\footnotesize
\caption{Workload-dependent SMC keys}
\label{tab:smc keys}

\begin{tabular}{l|lll|lll}
   & \multicolumn{3}{c|}{Mac Mini M1} &  \multicolumn{3}{c}{Mac Air M2}                                                     \\ \hline
    SMC keys & PDTR, & PHPC, & PHPS & PDTR, & PHPC, & PHPS \\
            & PMVR, & PPMR & PSTR & PMVC, & PSTR \\
    \hline

\end{tabular}%
\end{table}

\subsection{Threat model}
In this work, we assume the attacker is a user mode software and the victim is a kernel mode driver holding a secret. The victim may provide services to user mode software. The attacker might use these services to call/invoke kernel routines that operate on the secret data. The attacker, as a user mode application, has no direct access to the secret which belongs to the kernel mode driver. We assume the kernel mode driver implementation follows secure coding guidelines so it is not vulnerable to traditional timing side-channel attacks.

\subsection{Identifying power-correlated SMC keys}
Since official SMC key definitions for Apple M1/M2 systems are not publicly available, the initial phase of the research involved identifying the specific SMC keys that correlate with power consumption among the extensive collection of available keys. Based on the established naming convention adopted in x86-based Mac systems, it has been observed that SMC keys associated to power-related functionalities commonly start with an initial capital letter ``P'' \cite{powerSMCKeys}. By leveraging this information, we were able to significantly narrow down the pool of candidate keys to approximately 30.

In the next step, we used an open source tool \textit{smc-fuzzer} \cite{smc-fuzzer} to enumerate values of the candidate keys when the system is idle and busy, respectively.
For all the SMC keys on the Apple M2 system that start with the letter ``P'', we collected their corresponding values when the system is in an idle state, as well as when a stressor \textit{stress-ng} \cite{king2017stress} workload, which performs matrix operations on all available cores, is running. These two scenarios are chosen because they are expected to show large power difference.
By conducting a side-by-side comparison between the results under both scenarios, we identified several SMC keys that exhibit large variations in their values. Those keys are considered correlated with power and are selected as the candidates for the next step analysis.
The same experiment was also replicated on Apple M1. Table \ref{tab:smc keys} summarizes the list of  SMC keys identified on both systems, respectively.

\begin{table*}[]
\caption{TVLA analysis on Apple M2 system for selected SMC keys between different plaintexts for the user space AES workload}
\label{tab:TVLA_m2}
\resizebox{\textwidth}{!}{%
\begin{tabular}{l|lll|lll|lll|lll|lll}
SMC key       & \multicolumn{3}{c|}{PHPC}                                                                   & \multicolumn{3}{c|}{PDTR}                                                                   & \multicolumn{3}{c|}{PHPS}                                                                 & \multicolumn{3}{c|}{PMVC}                                                                   & \multicolumn{3}{c}{PSTR}                                                                    \\ \hline
Plaintext & All 0s                        & All 1s                       & Random                       & All 0s                        & All 1s                       & Random                       & All 0s                       & All 1s                      & Random                       & All 0s                        & All 1s                       & Random                       & All 0s                       & All 1s                        & Random                       \\ \hline
All 0s'   & -0.18                         & {\color[HTML]{FE0000} 20.94} & {\color[HTML]{FE0000} 11.49} & {\color[HTML]{3531FF} 8.73}   & {\color[HTML]{FE0000} 29.58} & {\color[HTML]{FE0000} 25.05} & 0.87                         & {\color[HTML]{009901} 2.14} & {\color[HTML]{009901} 2.02}  & {\color[HTML]{3531FF} 9.49}   & {\color[HTML]{FE0000} 32.16} & {\color[HTML]{FE0000} 27.33} & {\color[HTML]{3531FF} 21.30} & {\color[HTML]{FE0000} 12.96}  & {\color[HTML]{FE0000} 27.41} \\
All 1s'   & {\color[HTML]{FE0000} -21.09} & 0.09                         & {\color[HTML]{FE0000} -8.87} & {\color[HTML]{FE0000} -20.42} & 0.02                         & {\color[HTML]{009901} -4.49} & {\color[HTML]{009901} -1.97} & -0.69                       & {\color[HTML]{009901} -0.84} & {\color[HTML]{FE0000} -22.45} & 0.09                         & {\color[HTML]{FE0000} -5.15} & {\color[HTML]{FE0000} 9.13}  & 0.37                          & {\color[HTML]{FE0000} 15.16} \\
Random'   & {\color[HTML]{FE0000} -11.60} & {\color[HTML]{FE0000} 9.28}  & 0.43                         & {\color[HTML]{FE0000} -15.28} & {\color[HTML]{FE0000} 5.01}  & 0.55                         & {\color[HTML]{009901} -0.53} & {\color[HTML]{009901} 0.74} & 0.61                         & {\color[HTML]{FE0000} -17.54} & {\color[HTML]{FE0000} 4.92}  & -0.23                        & {\color[HTML]{FE0000} 5.99}  & {\color[HTML]{FE0000} -15.03} & -0.24                       
\end{tabular}%
}
\vspace{-10pt}
\end{table*}

\label{sub:t-test}
\subsection{Identifying data-dependent SMC keys}
Our next step is to identify the data-dependent SMC keys from the previous list of power-correlated keys. To accomplish this, we implemented a Proof-of-Concept (PoC) that executes a workload which accepts and processes distinct input data for a fixed number of iterations. The workload is constant-cycle to avoid impact from timing side-channel. Values of all the selected SMC keys are measured and logged in a \textit{trace} after the workload finishes. The process is repeated to collect multiple \textit{traces} containing the SMC key values. 

In order to make the potential data-dependency more observable, we opted to replicate the workload and execute it simultaneously on three P-cores. The three copies execute the same workload using the same input data at any time, therefore the data-dependent power consumption is amplified. We did not run additional copy on the fourth performance cores (P-core) or the efficiency cores (E-cores) 
We applied the Test Vector Leakage Assessment (TVLA) \cite{gilbert2011testing} to validate if any of the pre-selected SMC key value traces show data-dependency. TVLA utilizes Welch's t-test to assess side-channel leakage of cryptographic implementations. A Welch's t-test compares two datasets, A and B, by computing a statistic score, the \textit{t-score}. A $|\textit{t-score}|>4.5$ indicates that the two datasets are statistically distinguishable with 99.999\% confidence.
For this experiment, we selected the implementation of AES-128 encryption from \textit{AES-Intrinsics} \cite{AES-Intrinsics} as the testing workload, which utilizes the ARMv8 equivalent to the cryptographic extension AES instructions (e.g., \textit{AESE} and \textit{AESMC}). We observed that the SMC key values are updated approximately every one second. Therefore, in the workload we repeated the AES encryption of the same plaintext so it takes slightly longer than a second. We collected 10k SMC key value traces corresponding to the encryption of each of the three chosen plaintexts - (\textit{All\_0s}, \textit{All\_1s}, and \textit{Random}) - with a fixed key. We then applied TVLA between every possible pair of the chosen plaintexts. Table \ref{tab:TVLA_m2} shows the calculated \textit{t-score} on the Apple M2 system, where different colors mean:
\vspace{-2pt}
\begin{itemize}
    \item \textcolor{red}{True positive}: two traces with different plaintexts are distinguishable ($|\textit{t-score}|\geq4.5$).
    \item \textcolor{black}{True negative}: two traces with the same plaintext are non-distinguishable ($|\textit{t-score}|<4.5$).
    \item \textcolor{blue}{False positive}: two traces with the same plaintext are distinguishable ($|\textit{t-score}|\geq4.5$).
    \item \textcolor{green}{False negative}: two traces with different plaintexts are non-distinguishable ($|\textit{t-score}|<4.5$).
\end{itemize}
If a SMC key is data-dependent, we expect it shows more true positive/negative and less false positive/negative. Out of the selected SMC keys, one key, namely \textit{PHPC}, stands out by demonstrating true positive and true negative results consistently. Remarkably, \textit{PHPC} exhibits no instances of false positive or false negative correlations. This finding strongly suggests that \textit{PHPC} has the most robust and reliable correlation with the data.
In the case of \textit{PDTR}, \textit{PMVC}, and \textit{PSTR}, these SMC keys exhibit a combination of true positive and true negative results for a majority of the pairs. However, they also display several instances of false positive or false negative. This indicates a somewhat weaker data-correlation compared to \textit{PHPC}. Conversely, the SMC key \textit{PHPS} primarily generates false negative correlations for most of the pairs and yields no true positive correlations, suggesting a limited or no data-correlation.
As a result of our analysis, we have confirmed the data-dependency for all the selected SMC keys except for \textit{PHPS} on Apple M2 system. Additionally, we conducted TVLA on the collected traces of the \textit{PHPC} key values collected on the Apple M1 platform, affirming a similar data-dependency pattern for the \textit{PHPC} key on that system as well.

\label{sub:aes-key-recovery-poc}
\subsection{AES encryption key extraction}
Following the confirmation of data-dependency in the SMC key values, we then assessed the feasibility of extracting secrets from a victim program. In a manner similar to the previous experiment, we used a victim program performing AES-128 encryption repeatedly, which carries out encryption operations using a secret key that remains inaccessible to the attacker. We still assume the attacker can perform \textit{know-plaintext attack}, in which he/she controls the plaintext. In contrast to providing fixed plaintexts like the TVLA analysis, the attacker now provides random plaintext inputs into the victim program.
During this process, the attacker records the plaintext, the generated ciphertext, and the corresponding SMC key values right after the encryption operation. As a starting point, the AES-128 encryption takes place in user mode of this PoC for the purpose to minimize noise. We will discuss extracting kernel secret later.

In this experiment, we collected 1 million SMC key value traces on the M2 system and 350 thousand traces on the M1 system, respectively. We then performed Correlation Power Analysis (CPA) to analyze the collected traces. CPA involves computing correlations between the power side-channel traces and a hypothetical power model derived from the Hamming Weight (HW) or Hamming Distance (HD) of intermediate states. In our CPA approach, the power traces used for analysis correspond to the SMC key value traces we collected. The hypothetical power models employed are similar to those used in traditional CPA. These models incorporate the HW or HD of intermediate states, which are:

\begin{itemize}
    \item \textbf{Rd0-HW}: HW after the first \textit{AddRoundKey} operation to recover initial round key
    \item \textbf{Rd10-HW}: HW before the last round \textit{SubBytes} operation to recover round \#10 key
    \item \textbf{Rd10-HD}: HD between last round input and ciphertext to recover round \#10 key. 
\end{itemize}

For each targeted key byte, the hypothetical power model is constructed using HW/HD of intermediate round state for all possible key guesses. We calculated the correlation coefficient between the SMC key value trace and the hypothetical power model, then ranked all key guesses in descending order based on correlation coefficient. The outcome of the CPA test is the rank of the correct key byte among all key guesses, with a value of 1 indicating successful recovery of the secret key byte. The average rank of all correct key bytes is measured by \emph{Guessing Entropy} (GE). Lower GE indicates lower ranks in general across all key bytes and $GE=0$ indicates recovery of all key bytes.

\begin{table}[]
\footnotesize
\centering
\caption{Rank of each AES key byte applying CPA on collected SMC key value traces with Round 0 HW power model on Macbook Air M2 and M1 (PHPC only) }
\label{tab:cpa_compare_keys}
\begin{tabular}{c|ccccc}
   \#key byte& \multicolumn{1}{l}{PHPC}          & \multicolumn{1}{l}{PDTR}          & \multicolumn{1}{l}{PMVC}          & \multicolumn{1}{l}{PSTR}  & \multicolumn{1}{l}{PHPC (M1)} \\ \hline
0  & {\color[HTML]{FFC702} \textbf{7}} & {\color[HTML]{FF0000} \textbf{1}} & {\color[HTML]{FFC702} \textbf{3}} & 211                       & {\color[HTML]{FFC702} \textbf{9}}\\
1  & {\color[HTML]{FFC702} \textbf{7}} & {\color[HTML]{FFC702} \textbf{7}} & {\color[HTML]{FFC702} \textbf{3}} & 22                        & 19\\
2  & {\color[HTML]{FF0000} \textbf{1}} & {\color[HTML]{FFC702} \textbf{5}} & {\color[HTML]{FFC702} \textbf{3}} & 188                       & {\color[HTML]{FFC702} \textbf{4}}\\
3  & 11                                & 11                                & 12                                & 189                       & 12\\
4  & {\color[HTML]{FFC702} \textbf{5}} & {\color[HTML]{FF0000} \textbf{1}} & {\color[HTML]{FF0000} \textbf{1}} & 151                       & {\color[HTML]{FF0000} \textbf{1}}\\
5  & {\color[HTML]{FFC702} \textbf{4}} & 15                                & 12                                & 223                       & 31\\
6  & {\color[HTML]{FFC702} \textbf{4}} & {\color[HTML]{FFC702} \textbf{6}} & 14                                & 113                       & 16\\
7  & 13                                & {\color[HTML]{FFC702} \textbf{8}} & 12                                & 39                        & {\color[HTML]{FFC702} \textbf{5}}\\
8  & {\color[HTML]{FF0000} \textbf{1}} & 15                                & 17                                & 201                       & {\color[HTML]{FFC702} \textbf{9}}\\
9  & 37                                & 16                                & 22                                & 101                       & 18\\
10 & {\color[HTML]{FF0000} \textbf{1}} & {\color[HTML]{FFC702} \textbf{5}} & 11                                & 214                       & {\color[HTML]{FFC702} \textbf{7}}\\
11 & {\color[HTML]{FF0000} \textbf{1}} & {\color[HTML]{FFC702} \textbf{2}} & {\color[HTML]{FFC702} \textbf{2}} & 117                       & {\color[HTML]{FFC702} \textbf{2}}\\
12 & {\color[HTML]{FF0000} \textbf{1}} & {\color[HTML]{FFC702} \textbf{2}} & {\color[HTML]{FF0000} \textbf{1}} & 146                       & {\color[HTML]{FF0000} \textbf{1}}\\
13 & {\color[HTML]{FFC702} \textbf{4}} & 12                                & 13                                & 184                       & 36\\
14 & {\color[HTML]{FF0000} \textbf{1}} & {\color[HTML]{FFC702} \textbf{9}} & {\color[HTML]{FFC702} \textbf{8}} & 18                        & 25\\
15 & 26                                & 24                                & 14                                & 137                       & 50\\ \hline
GE & 31.0          & 41.6          & 42.8          & 109.3  & 40.9
\end{tabular}
\vspace{-10pt}
\end{table}

Table \ref{tab:cpa_compare_keys} presents the rank of each secret key byte and the \emph{Guessing Entropy} (GE) using the \textit{Rd0-HW} power model on M1/M2 systems. Key bytes with a rank of 1, indicating successful recovery, are highlighted in red. Those with ranks below 10, suggesting near recovery, are in yellow.
On the Macbook Air M2, CPA analysis of \emph{PHPC} traces successfully recovers 6 key bytes and nearly recovers another 6. For \emph{PDTR} and \emph{PMVC} traces, CPA yields low ranks for most key bytes (10/16 for \emph{PDTR}, 7/16 for \emph{PMVC}). However, CPA analysis of \emph{PSTR} traces fails to recover any key bytes. On the Apple M1 mini, CPA testing of the \emph{PHPC} key successfully recovers 2 bytes and nearly recovers 6 others, consistent with prior TVLA findings.

\begin{figure}
    \centering
    \begin{tabular}{@{}c@{}}
        \includegraphics[width=0.9\linewidth]{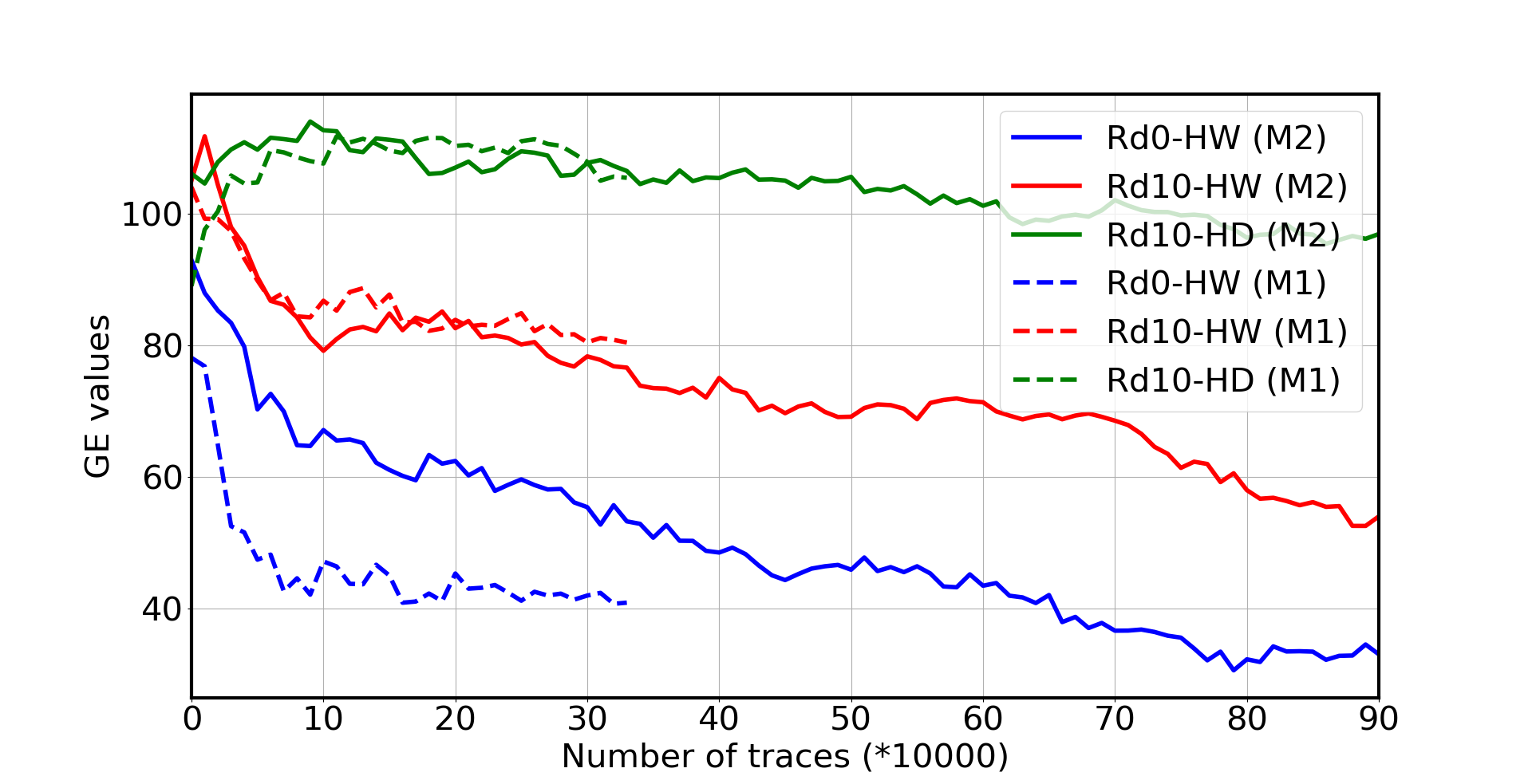} \\
        \\[-\abovecaptionskip]
        \small (a)
    \label{fig:GE_PHPC_M2}
    \end{tabular}
    \begin{tabular}{@{}c@{}}
        \includegraphics[width=0.9\linewidth]{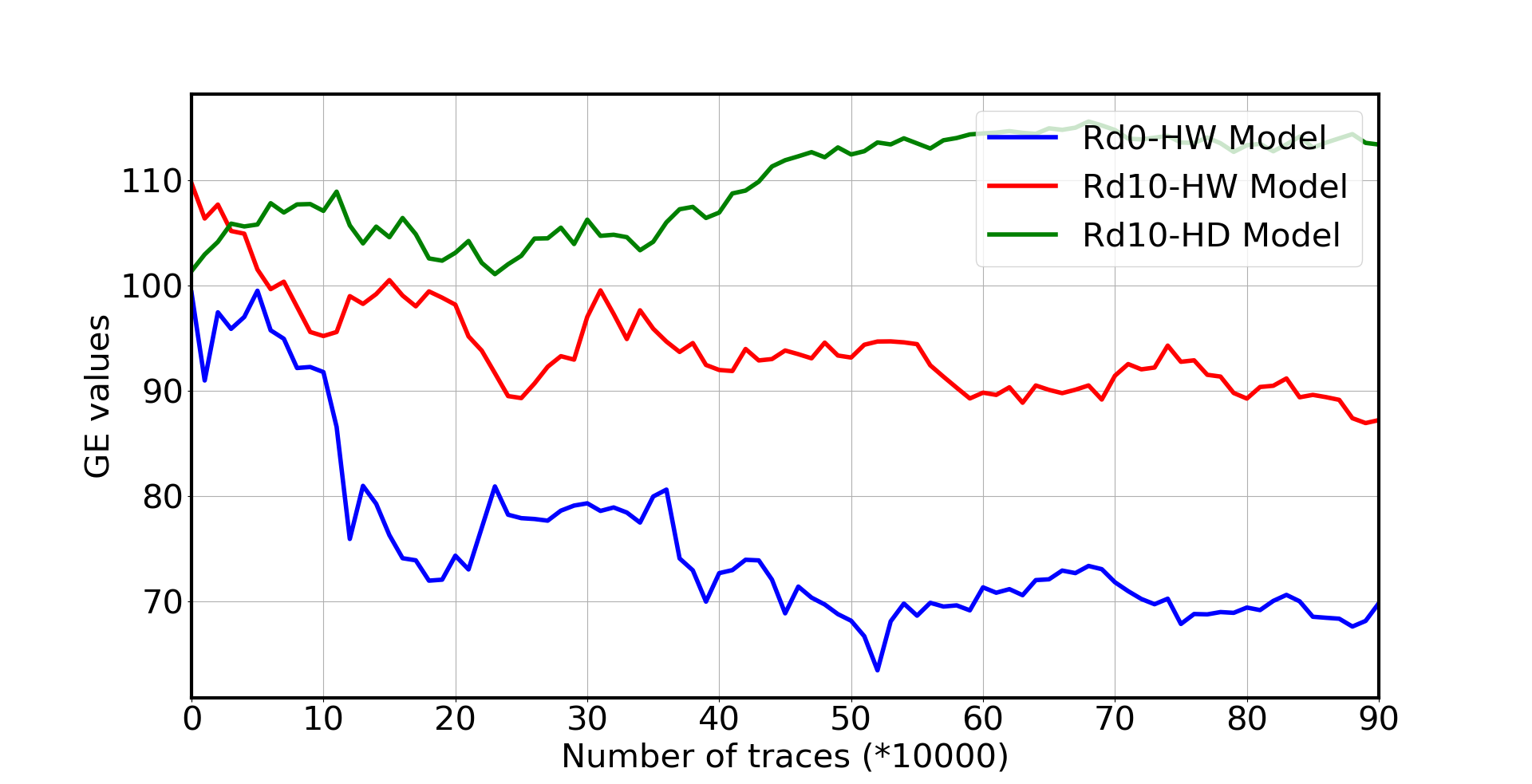} \\
        \\[-\abovecaptionskip]
        \small (b)
    \label{fig:GE_PHPC_M2_Kext}
    \end{tabular}
    \caption{GE trend against collected \emph{PHPC} SMC key value traces for CPA targeting (a) user space AES encryption on Apple M1 Mini and M2 Air systems, and (b) AES kernel module on the Macbook Air M2 system}\label{fig:GE}
    \vspace{-6pt} 
\end{figure}

Figure \ref{fig:GE}(a) illustrates how the GE varies with the number of \emph{PHPC} key value traces and different power models for both M1 and M2 systems. The GE trend for the M1 system is shorter due to a smaller dataset of \emph{PHPC} traces (350K traces). The trends suggest that accumulating more traces improves the likelihood of recovering all key bytes.
Different power models show varying convergence patterns. The \emph{Rd0-HW} model converges quickest, highlighting its superior key recovery capability. The \emph{Rd10-HW} model converges as well, but more slowly. In contrast, the \emph{Rd10-HD} model shows little convergence, indicating it is less effective for key recovery.
These findings affirm the \emph{Rd0-HW} model's effectiveness and dependability for extracting information from SMC key value traces on Apple's M1/M2 platforms.

\begin{table*}[]
\caption{TVLA analysis on selected SMC key traces corresponding to encryption of different plaintext inputs to AES kernel module on the Macbook Air M2 system }
\label{tab:TVLA_m2_kernel}
\resizebox{\textwidth}{!}{%
\begin{tabular}{l|lll|lll|lll|lll|lll}
SMC key       & \multicolumn{3}{c|}{PHPC}                                                                   & \multicolumn{3}{c|}{PDTR}                                                                   & \multicolumn{3}{c|}{PHPS}                                                                 & \multicolumn{3}{c|}{PMVC}                                                                   & \multicolumn{3}{c}{PSTR}                                                                    \\ \hline
Plaintext & All 0s                        & All 1s                       & Random                       & All 0s                        & All 1s                       & Random                       & All 0s                       & All 1s                      & Random                       & All 0s                        & All 1s                       & Random                       & All 0s                       & All 1s                        & Random                       \\ \hline
All 0s'   & 2.78                         & {\color[HTML]{FE0000} 19.28} & {\color[HTML]{FE0000} 9.41} & {\color[HTML]{3531FF} 13.84}   & {\color[HTML]{FE0000} 41.52} & {\color[HTML]{FE0000} 43.01} & 2.72                         & {\color[HTML]{009901} 3.60} & {\color[HTML]{FE0000} 6.51}  & {\color[HTML]{3531FF} 15.13}   & {\color[HTML]{FE0000} 45.38} & {\color[HTML]{FE0000} 47.10} & {\color[HTML]{3531FF} 40.66} & {\color[HTML]{FE0000} 18.45}  & {\color[HTML]{FE0000} 37.50} \\
All 1s'   & {\color[HTML]{FE0000} -17.91} & -0.76                         & {\color[HTML]{FE0000} -11.12} & {\color[HTML]{FE0000} -30.27} & -2.16                         & {\color[HTML]{009901} -0.26} & {\color[HTML]{009901} -3.99} & -3.12                       & {\color[HTML]{009901} -0.11} & {\color[HTML]{FE0000} -32.93} & -2.44                         & {\color[HTML]{009901}-0.48} & {\color[HTML]{FE0000} -18.45}  & 1.66                          & {\color[HTML]{FE0000} 20.01} \\
Random'   & {\color[HTML]{FE0000} -6.77} & {\color[HTML]{FE0000} 10.14}  & -0.04                         & {\color[HTML]{FE0000} -30.84} & {\color[HTML]{FE0000} -0.73}  & -0.99                         & {\color[HTML]{009901} -3.85} & {\color[HTML]{009901} 0.11} & 0.03                         & {\color[HTML]{FE0000} -33.30} & {\color[HTML]{009901}-0.56}  & -1.03                        & {\color[HTML]{009901} 0.73}  & {\color[HTML]{FE0000} -20.70} & -2.16                       
\end{tabular}%
}
\end{table*}

\subsection{Targeting a kernel module}
To assess the attack's viability across different security levels, we implemented a kernel module on the M2 system to act as the target program, with the attacker being a user space application that has read access to the SMC key values via the \texttt{IOKit} framework's interface. This victim kernel module functions as an encryption service, taking plaintext from a user application, performing encryption repeatedly for a set number of cycles, and then storing the resulting ciphertext in a buffer accessible to the user application. In our setup, a single P-core thread triggers the device driver to encrypt the plaintext.

We applied TVLA to the SMC key traces collected to assess data dependency, as outlined in section \ref{sub:aes-key-recovery-poc}. We recorded SMC key traces while encrypting different plaintext inputs using the device driver. The \textit{t-scores} for these traces during fixed-key encryption on the MacBook Air M2 are displayed in Table \ref{tab:TVLA_m2_kernel}, with color-coding for TVLA results following the standards set in section \ref{sub:t-test}. The data dependency patterns observed were consistent with those from a user space victim, previously discussed. The \textit{PHPC} key traces demonstrated the most significant data dependency, with \textit{PDTR}, \textit{PMVC}, and \textit{PSTR} also showing dependency, and \textit{PHPS} traces revealing the least correlation.

We further conducted a CPA test on the SMC keys whose values exhibit data correlation in the TVLA test. The objective is to recover the AES encryption key from the device driver. We collected one million traces for each of the SMC keys and then apply CPA with the hypothetical power models considered in section \ref{sub:aes-key-recovery-poc} to obtain the GE metrics. 
Fig. \ref{fig:GE}(b) illustrates the GE trend with different power models against the one million traces collected for \textit{PHPC} SMC key. We observed a converging trend indicating a reduction in the rank of correct key bytes with increasing number of collected traces. Among the power models tested, the \textit{Rd0-HW} power model demonstrates the strongest correlation indicated by the fastest convergence of the GE metric, while the \textit{Rd10-HD} power model do not exhibit any convergence.
Additionally, we observed that the GE metric in Fig. \ref{fig:GE}(b) converges approximately two times slower compared to the user mode AES application (Fig. \ref{fig:GE}(b)). This can be attributed to the decreased \textit{Signal-to-Noise Ratio (SNR)} of the collected traces due to system call invocations, as well as the lower number of victim threads in the kernel module test. The experimental findings from the CPA test indicate that an unprivileged attacker can compromise the confidentiality of assets protected by the kernel using SMC key as a side-channel.

\begin{table}
\caption{TVLA analysis on the `PCPU' channel traces (from IOReport) and execution time traces (during \texttt{lowpowermode} throttling)  on Macbook Air M2 system}
\label{tab:TVLA_m2_throttling}
\resizebox{\columnwidth}{!}{%
\begin{tabular}{l|lll|lll}
       & \multicolumn{3}{c|}{PCPU (IOReport)} &  \multicolumn{3}{c}{Time (during throttling)}                                                     \\ \hline

Plaintext & All 0s & All 1s & Random & All 0s & All 1s & Random \\ \hline
All 0s'   & -0.34 & {\color[HTML]{009901} -1.41} & {\color[HTML]{009901} -2.68} & -0.34 & {\color[HTML]{009901} -1.07} & {\color[HTML]{009901} -0.29} \\
All 1s'   & {\color[HTML]{009901} 2.08} & 0.28                         & {\color[HTML]{009901} -1.95} & {\color[HTML]{009901} 0.84} & 0.10                         & {\color[HTML]{009901} 0.90} \\
Random'   & {\color[HTML]{009901} -2.68} & {\color[HTML]{009901} 1.95}  & -1.09 & {\color[HTML]{009901} -0.41} & {\color[HTML]{009901} -1.13}  & -0.36  \\         
\end{tabular}%
}
\end{table}

\subsection {IOReport Energy Reporting}
We have also examined alternative power reporting interfaces beside the SMC keys. 
We leveraged an open-source tool \texttt{socpowerbud} \cite{socpowerbud} which utilizes \texttt{IOReport} functions imported from \emph{Foundation.h} header file to dump system level statistics for user space software to read. These statistics are further categorized into multiple group, where each group has multiple channels. Among all the groups, we find that ``PCPU'' channel within ``Energy Model'' group reports energy consumption of the P-cores, which is potentially another exploitable power side-channel\footnote{Please note that ``PCPU'' is not a SMC key, although it aligns with SMC key's naming convention.}. We assess the side-channel leakage from the collected ``PCPU’’ channel readings by applying TVLA to check if it exhibits data-dependency. The result (shown at Table \ref{tab:TVLA_m2_throttling} first column), however, indicates no data correlation as there are all false negatives between different plaintext pairs. We attribute the observation to two reasons. first, ``PCPU’’ channel reports energy with resolution of mJ, so that the averaged power has a coarser-grained resolution (mW) compared to that derived from SMC keys (uW); more importantly, we suspect ``Energy Model’’ group reports estimated energy model calculated based on core utilization instead of actual sensor reading, therefore doesn't reveal data-dependency.

%% file: sections/throttling.tex
We also explored the viability of conducting frequency throttling side-channel attacks on the Apple M2 system, another variant of software-based power side-channel attacks previously identified in x86 platforms. This method requires the system to hit a reactive power limit (e.g., power limit) while the victim program is operational. To trigger this, we ran multiple instances of the victim program across various cores, augmenting with stressor code on the other cores where necessary to exceed the system's power threshold. However, with such settings we observed the system's thermal limit was consistently reached before any power-based throttling. While reaching thermal limit also triggers frequency throttling and previous work \cite{taneja2023hot} shows it reveals certain level of side-channel information, we found that its SNR is too low to mount any realistic attack.

\noindent\textbf{Finding Reactive Power Limits}: 
The observations outlined above prompted us to explore strategies for reaching the power limit without triggering the thermal limit. One potential method involves lowering the default power limit. However, we encountered a lack of publicly accessible documentation regarding power management and specific reactive limits for the M2 system. Our investigation of modifiable System Management Controller (SMC) keys, through visual inspection and the use of the \textit{smc-fuzzer} tool \cite{smc-fuzzer}, did not reveal any keys related to reactive limit configurations.

Further analysis using \texttt{pmset}, a utility provided by Apple's macOS for adjusting power management settings, led us to discover a tunable binary setting named \texttt{lowpowermode}. Activating \texttt{lowpowermode} by setting it to 1, we conducted tests to observe the effects on CPU power and core frequencies under varying workload conditions. Starting with a single AES-128 encryption thread on one P-core, we incrementally increased the load by adding identical threads across more cores. We noted that the P-cores maintained a consistent frequency of 1.968 GHz when the total CPU power consumption remained under 4W. Upon attempting to increase the workload further, we observed a power cap at 4W, coinciding with the onset of frequency throttling in the P-cores. Given that the CPU temperature remained low during this throttling, we ruled out thermal factors as the cause and concluded that a power limit of 4W is enforced in \texttt{lowpowermode}, with P-core throttling occurring upon reaching this threshold. Interestingly, the efficiency cores (E-cores) did not exhibit throttling and continued to operate at a stable frequency of 2.424 GHz.

\noindent\textbf{Triggering Frequency Throttling}: Upon identifying a method to induce power limit-based throttling, we aimed to harness this for a potential side-channel attack. Two prerequisites were established: 1) secret-dependent computations must be confined to the P-cores, as the E-cores do not experience throttling, and 2) the CPU's power consumption must surpass the 4W threshold set by the \texttt{lowpowermode}. Drawing from prior experiments, we chose an AES-128 encryption routine from AES-Intrinsics \cite{AES-Intrinsics} as our target. In \texttt{lowpowermode}, running the AES-128 workload on all four P-cores resulted in a power draw of only 2.8W, which is insufficient to induce throttling. Consequently, we introduced additional stressors on the E-cores to elevate the system's power usage.
The challenge was to confine the AES threads to the P-cores while running stressors on the E-cores. We addressed this by adjusting the scheduler policy to round-robin (\texttt{SCHED\_RR}) and setting the AES threads to the highest priority, guiding the operating system's scheduler to favor the P-cores for these threads. Our stressor code executed floating-point multiplication (\textit{fmul}) instructions between two constant operands, creating a steady and secret-independent power load that increased the system's total power consumption without introducing power fluctuations. Our experiments showed that by operating four AES threads on the P-cores alongside a \textit{fmul} stressor on the E-cores, we reached the 4W power ceiling, triggering throttling on the P-cores.

\noindent\textbf{Experimental Results}: Adopting the setup outlined earlier, we replaced SMC key measurements with execution time the AES threads, as detailed in Section \ref{sub:t-test}. We conducted TVLA analysis on the timing traeces with results presented in the second column of Table \ref{tab:TVLA_m2_throttling}. The absence of true positives suggests that the timing (and throttled frequency) is not data-dependent. Notably, the "PHPS" SMC key value also peaked at 4W during throttling. Given the lack of data-dependency in "PHPS" values (Table \ref{tab:TVLA_m2}), we infer that \texttt{lowpowermode}'s frequency throttling may rely on "PHPS" rather than actual power use, explaining the lack of data correlation.

%% file: sections/mitigation.tex
To address the vulnerabilities highlighted by the PLATYPUS attack~\cite{Lipp2020Platypus}, both Intel and AMD have taken measures to remove the user space software's access to the RAPL (Running Average Power Limit) energy counters from the Linux kernel drivers~\cite{INTEL-SA-00389,CVE-2020-12912}. Additionally, Intel has introduced a RAPL filtering mechanism, which blends random energy noise with the energy reporting and adjusts the update interval, aiming to lower the SNR of power consumption measurements and hinder the extraction of sensitive data \cite{INTEL-RAPL-GUIDANCE}.

In light of our research, we propose that analogous countermeasures could be effective for the SMC key power side-channel vulnerabilities we've identified. Restricting user space access to power-related SMC keys would prevent user space attackers from exploiting these values. Incorporating noise into power consumption readings could further degrade the SNR, diminishing the threat posed by these side-channel attacks.
As of our publication, there is no indication that Apple has implemented specific mitigation against the vulnerabilities we've discussed. It is imperative for system vendors, including Apple, to consider the security implications of power side-channel attacks and to establish robust defenses to secure sensitive information.

Our study contributes to a growing amount of work that demonstrates the risk of software-based power side-channel attacks across different CPU architectures. These findings highlight an industry-wide issue, stressing the importance for all vendors to acknowledge the threat and to actively seek out architectural solutions and preventative measures to counteract emerging software-based power side-channel attacks.

%% file: sections/conclusion.tex
This paper has explored the vulnerabilities arising from software-based power side-channel attacks and their implications across CPU architectures. Our experiments and analysis have revealed the data dependency of the power consumption reported by the selected System Management Controller (SMC) key values on ARM-based Apple Silicon M1/M2 systems. Furthermore, we demonstrated that a user space attacker can exploit the unprivileged access to the SMC key values to extract secrets from the kernel device driver. Our research underscores the need for industry-wide awareness and proactive exploration of architectural solutions to prevent software-based power side-channel attacks. 